\documentclass[aps,nofootinbib,floatfix,preprint]{revtex4}
\usepackage{epsfig}
\usepackage{graphicx}
\usepackage{multirow}
\begin{document} 
\title{\textbf{$T_{bbb}$: a three $B-$meson bound state}}
\author{H.~Garcilazo} 
\email{humberto@esfm.ipn.mx} 
\affiliation{Escuela Superior de F\' \i sica y Matem\'aticas, \\ 
Instituto Polit\'ecnico Nacional, Edificio 9, 
07738 M\'exico D.F., Mexico} 
\author{A.~Valcarce} 
\email{valcarce@usal.es} 
\affiliation{Departamento de F\'\i sica Fundamental and IUFFyM,\\ 
Universidad de Salamanca, E-37008 Salamanca, Spain}
\date{\today} 
\begin{abstract}
By solving exactly the Faddeev equations for the bound-state problem of three
mesons, we demonstrate that current theoretical predictions pointing to 
the existence of a deeply-bound doubly bottom axial vector tetraquark lead to the 
existence of a unique bound state of three $B$ mesons.
We find that the $BB^*B^*-B^*B^*B^*$ state with quantum numbers
$(I)J^P=(1/2)2^-$, $T_{bbb}$, is about 90\,MeV below any possible three $B$-meson threshold
for the reported binding of the doubly bottom axial vector tetraquark, $T_{bb}$.
\end{abstract}
\maketitle 

\section{Introduction.} 
There is a broad theoretical consensus about the existence of a 
deeply-bound doubly bottom tetraquark with quantum numbers $(i)j^p=(0)1^+$ 
strong- and electromagnetic-interaction 
stable~\cite{Fra17,Kar17,Eic17,Bic16,Vij09,Ric18,Luo17,Duc13,Cza18,Ade82}. 
In the pioneering work of Ref.~\cite{Ade82} it was shown that
$QQ\bar q \bar q$ four-quark configurations become more and more
bound when the mass ratio $M_Q/m_q$ increases, the critical value 
for binding being somewhat model dependent.

Lattice QCD calculations find unambiguous signals for a stable $j^p=1^+$ 
bottom-light tetraquark~\cite{Fra17}. Based on a diquark hypothesis, Ref.~\cite{Kar17} 
uses the discovery of the $\Xi_{cc}^{++}$ baryon~\cite{Aai17} 
to {\em calibrate} the binding energy in a $QQ$ diquark. Assuming that
the same relation is true for the $bb$ binding energy in a tetraquark,
it concludes that the axial vector $bb\bar u \bar d$ state is stable.
The Heavy-Quark Symmetry analysis of Ref.~\cite{Eic17} predicts the existence
of narrow doubly heavy tetraquarks. Using as input for the doubly bottom baryons, not 
yet experimentally measured, the diquark-model calculations
of Ref.~\cite{Kar17} also leads to a bound axial vector $bb\bar u \bar d$ 
tetraquark. Other approaches, using Wilson twisted mass lattice QCD~\cite{Bic16}, 
also find a bound state. Few-body calculations using quark-quark 
Cornell-like interactions~\cite{Vij09,Ric18},
simple color magnetic models~\cite{Luo17}, QCD sum rule analysis~\cite{Duc13}, or
phenomenological studies~\cite{Cza18} come to similar conclusions.

The possible existence of deuteron-like hadronic molecular states 
made of vector-vector or pseudoscalar-vector two-meson systems
was proposed in Ref.~\cite{Tor91} in an 
exploratory study suggesting the {\it deusons}, two-meson
states bound by the one-pion exchange potential. This scenario of
meson-meson stable states bound by some interacting potential, has later 
on been frequently used to draw conclusions about the existence of 
hadronic molecules~\cite{Man93,Eri93,Clo10} (see
Refs.~\cite{Che16} for a recent compendium). The constituent quark and the
meson-meson approaches to hadronic molecules must be equivalent~\cite{Har81},
although, as will be discussed below, to get the results of the constituent quark 
approach would, in general, require
a coupled-channel meson-meson study~\cite{Via09}. 
  
It is also worth to emphasize that when a two-body interaction is attractive, if the 
two-body system is merged with nuclear matter and the Pauli principle does not impose 
severe restrictions, the attraction may be reinforced. We find the simplest example of the effect 
of additional particles in the two-nucleon system. The deuteron, $(i)j^p=(0)1^+$, 
is bound by $2.225$ MeV, while the triton, $(I)J^P=(1/2)1/2^+$, is bound by $8.480$ MeV, 
and the $\alpha$ particle, $(I)J^P=(0)0^+$, is bound by $28.295$ MeV. The binding per 
nucleon $B/A$ increases as $1:3:7$. Thus, a challenging question is if the existence
if a deeply bound two $B$-meson system~\footnote{The binding energy for the axial vector
doubly bottom tetraquark reported in Refs.~\cite{Fra17,Kar17,Eic17,Bic16,Vij09,Ric18,Luo17,Duc13,Cza18,Ade82}
ranges between 90 and 214 MeV.} could give rise to bound states of a larger number
of particles. As it was shown in Ref.~\cite{Gar17} the answer is by no means trivial,
because when the internal two-body thresholds of a three-body system are far away, they
conspire against the stability of the three-body system.

\section{Color dynamics.}
As it has been stated above, results based on meson-meson scattering or a constituent 
quark picture should be equivalent, provided that, in general, a coupled-channel
meson-meson approach would be necessary to reproduce the constituent quark picture~\cite{Har81,Via09}. 
To be a little more specific, let us note that
four-quark systems present a richer color structure than standard 
baryons or mesons. Although the color wave function for standard mesons 
and baryons leads to a single vector, working with four-quark states there 
are different vectors driving to a singlet color state out of colorless meson-meson ($\bf 1 \bf 1$) or 
colored two-body ($\bf{8} \bf{8}$, $\bf{\bar 3} \bf{3}$, or $\bf{6} \bf{\bar 6}$) components. Thus, dealing with
four-quark states an important question is whether one is in
front of a colorless meson-meson molecule or a compact state
(i.e., a system with two-body colored components). Note, however, that
any hidden color vector can be expanded as an infinite sum of colorless
singlet-singlet states~\cite{Har81}. This has been explicitly
done for compact $QQ\bar q \bar q$ states in Ref.~\cite{Via09}.

In the heavy-quark limit, the lowest lying tetraquark configuration
resembles the helium atom~\cite{Eic17}, a factorized system with separate dynamics
for the compact color $\bf \bar 3$ $QQ$ {\em nucleus} and for the
light quarks bound to the stationary color $\bf 3$ state, to construct
a $QQ\bar q \bar q$ color singlet. The validity of this argument 
has been mathematically proved and numerically
checked in Ref.~\cite{Via09}, see the probabilities shown in Table II
for the axial vector $bb\bar u \bar d$ tetraquark. It has been recently 
revised in Ref.~\cite{Ric18}, showing in Fig. 8 how  the probability of the 
$\bf 6 \bf \bar 6$ component in a compact $QQ\bar q\bar q$ tetraquark tends 
to zero for $M_Q \to \infty$.
Therefore, heavy-light compact bound states would be almost a pure $\bf{\bar 3} \bf{3}$ 
singlet color state and not a single colorless meson-meson $\bf 1 \bf 1$ molecule.
Such compact states with two-body colored components
can be expanded as the mixture of several physical meson-meson channels~\cite{Har81}, $BB^*$ and $B^*B^*$ 
for the axial vector $bb\bar u \bar d$ tetraquark (see Table II of Ref.~\cite{Via09})
and, thus, they can be also studied as an involved
coupled-channel problem of physical meson-meson states~\cite{Car12,Ike14}.

Our aim in this work is to solve exactly the Faddeev equations for the three-meson bound 
state problem using as input the two-body $t-$matrices of Refs.~\cite{Vij09,Via09,Gar17,Car12}, driving to 
the axial vector $bb\bar u \bar d$ bound state, $T_{bb}$, as an involved coupled-channel
system made of pseudoscalar-vector and vector-vector two $B$-meson components. We show that 
for any of the recently reported values of the $T_{bb}$ binding 
energy~\cite{Fra17,Kar17,Eic17,Bic16,Vij09,Ric18,Luo17,Duc13,Cza18,Ade82}, the three-body system 
$BB^*B^*-B^*B^*B^*$ with quantum numbers $(I)J^P=(1/2)2^-$, $T_{bbb}$, is between 43 to 90\,MeV
below the lowest three $B$-meson threshold. \\

\section{The three-body system.} 
\begin{table}[b!]
\caption{Different two-body channels $(i,j)$ contributing to the $(I)J^P=(1/2)2^-$
$B B^* B^* - B^* B^* B^*$ system.}
\begin{tabular}{|c|c|c|} 
\hline
Interacting pair & $(i,j)$ & Spectator  \\ 
\hline\hline
\multirow{2}{*}{$B B^*$}   & $(0,1)$   & \multirow{2}{*}{$B^*$} \\
                           & $(1,1)$   &                      \\ \hline
\multirow{2}{*}{$B^* B^*$} & $(0,1)$   & \multirow{2}{*}{$B^*$} \\
                           & $(1,2)$   &                      \\ \hline
$B^* B^*$                  & $(1,2)$   & $B$ \\ \hline																																					
\end{tabular}
\label{tab1}
\end{table}
\begin{figure}[t!]
\centering
\includegraphics[width=.75\columnwidth]{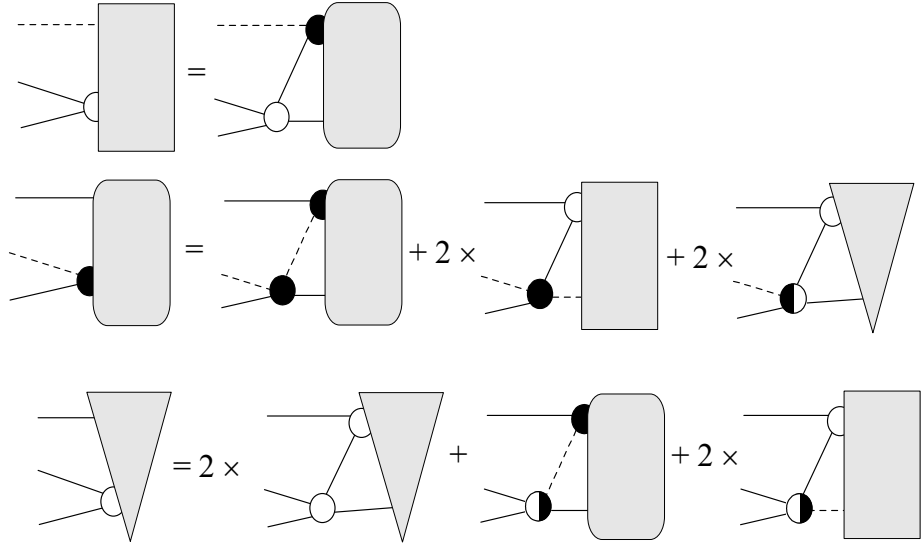}
\vspace*{-10.cm}
\caption{Diagramatic Faddeev equations for the three $B$-meson system.}
\label{fig1}
\end{figure}
Out of the possible spin-isospin three-body channels $(I)J^P$ made of $B$ and $B^*$ mesons,
we select those where, firstly, two-body subsystems containing two $B$-mesons are not allowed, 
because the $BB$ interaction does not show an attractive character; and, secondly, they contain 
the axial vector $(i)j^p=(0)1^+$ doubly bottom tetraquark, $T_{bb}$.
The three-body channel $(I)J^P=(1/2)2^-$ is the only one bringing together all these 
conditions to maximize the possible binding of the three-body system\footnote{Note
that the three-body channels with $J=0$ or $1$ would couple to two $B$-meson subsystems where 
no attraction has been reported~\cite{Fra17,Kar17,Eic17,Bic16,Vij09,Ric18,Luo17,Duc13,Cza18,Ade82},
whereas the $J=3$ would not contain a two-body subsystem with $j=1$, the quantum numbers
of the deeply bound doubly-bottom tetraquark. The same reasoning excludes the $I=3/2$ 
channels.}. 
We indicate in Table~\ref{tab1} 
the two-body channels contributing to this state that we examine in the following.

The Lippmann-Schwinger equation for the bound-state three-body problem is
\begin{equation}
T=(V_1+V_2+V_3)G_0 T \, ,
\label{eq1}
\end{equation}
where $V_i$ is the potential between particles $j$ and $k$ and $G_0$ is the
propagator of three free particles. The Faddeev decomposition of
Eq.~(\ref{eq1}),
\begin{equation}
T=T_1 + T_2 + T_3 \, ,   
\label{eq2}
\end{equation}
leads to the set of coupled equations,
\begin{equation}
T_i = V_i G_0 T\, .
\label{eq3}
\end{equation}
The Faddeev decomposition guarantees the uniqueness of the solution~\cite{Fad61}.
Eqs.~(\ref{eq3}) can be rewritten in the Faddeev form
\begin{equation}
T_i = t_i G _0(T_j + T_k) \, ,   
\label{eq4}
\end{equation}
with
\begin{equation}
t_i = V_i + V_i G_0 t_i \, ,
\label{eq5}
\end{equation}
where $t_i$ are the two-body $t-$matrices that already contain the coupling among 
all two-body channels contributing to a given three-body state, see Table~\ref{tab1}.
The two sets of equations~(\ref{eq3}) and~(\ref{eq4}) are completely equivalent for the
bound-state problem. In the case of two three-body systems that are coupled together, like $BB^*B^* - B^*B^*B^*$,
the amplitudes $T_i$  become two-component
vectors and the operators $V _i$, $t _i$, and $G _0$ become $2 \times 2$ matrices
and lead to the equations depicted in Fig. ~\ref{fig1}.
The solid lines represent the $B^*$
mesons and the dashed lines the $B$ meson. If in the second equation
depicted in Fig.~\ref{fig1} one drops the last term in the r.h.s. then the first
and second equations become the Faddeev equations of two identical
bosons plus a third one that is different~\cite{Gar17}. Similarly, if 
in the third equation depicted in Fig. 1 one drops the last two terms
this equation becomes the Faddeev equation of a system of three identical
bosons since in this case the three coupled Faddeev equations are all
identical~\cite{Gar17}. The additional terms in Fig.~\ref{fig1} are, of course,
those responsible for the coupling between the 
$BB^*B^*$ and $B^*B^*B^*$ components of the system. 
\begin{figure}[t!]
\centering
\includegraphics[width=0.85\columnwidth]{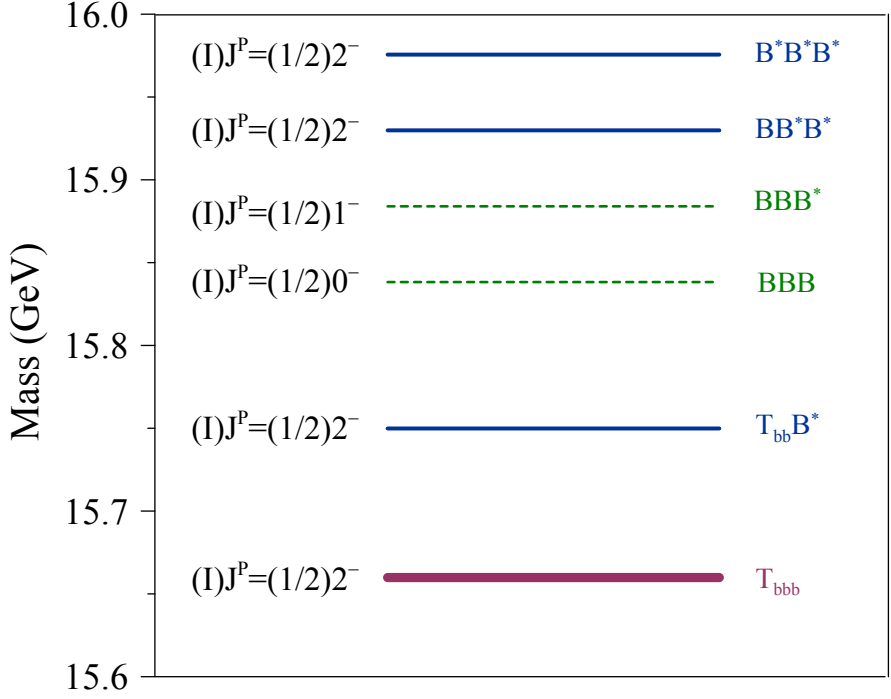}
\vspace*{-9.0cm}
\caption{Mass of the three-body $BB^*B^*-B^*B^*B^*$ bound-state $(I)J^P=(1/2)2^-$ 
$T_{bbb}$ (purple thick line),
compared to the different three $B$-meson strong (blue solid lines) and electromagnetic
decay thresholds (green dashed lines).}
\label{fig3}
\end{figure}

\section{Results.} 
We show in Fig.~\ref{fig3} the results of our calculation. The blue solid
lines stand for the different three $B$-meson strong decay thresholds of the
$BB^*B^*-B^*B^*B^*$ system with quantum numbers $(I)J^P=(1/2)2^-$, that we have denoted by $T_{bbb}$.
These thresholds are $B^*B^*B^*$, $BB^*B^*$ and $T_{bb}B^*$, where $T_{bb}$ represents the
axial vector $(i)j^p=(0)1^+$ doubly bottom tetraquark. The green dashed lines stand for the possible 
three-$B$ meson electromagnetic
decay thresholds, $BBB^*$ and $BBB$ with quantum number $(I)J^P=(1/2)1^-$ and $(I)J^P=(1/2)0^-$, respectively. 
Finally, the purple thick line indicates the energy of the $T_{bbb}$
state, that appears 90 MeV below the lowest threshold. The results shown in Fig.~\ref{fig1}
correspond to the binding energy of the $T_{bb}$ axial vector tetraquark obtained in Ref.~\cite{Fra17}. 

There is also a baryon-antibaryon threshold $\Omega_{bbb} -\bar p$ clearly decoupled
from the $T_{bbb}$, with a tetraquark-meson dominant component driving to the three $B$-meson bound
state, due to the orthogonality of the color wave function. 
The decay of the $T_{bbb}$ multiquark state $\left| \Psi_{T_{bbb}} \right\rangle$, with a dominant
tetraquark-meson color component\footnote{This is in contrast to the
analysis of Ref.~\cite{Gre84} where baryon-antibaryon annihilation into three-mesons
is studied by simple quark rearrangement.}, into a 
baryon ($B_1$) plus and antibaryon ($\bar B_2$) is forbidden if the transition amplitude 
$\left\langle B_1 \bar{B}_2  | { T} | \Psi_{T_{bbb}} \right\rangle$ vanishes. In principle
$T$ is the transition matrix (or $S$ matrix) which is
roughly $e^{iH}$, but since $\left| \Psi_{T_{bbb}} \right\rangle$ is a true eigenstate of $H$, 
the transition amplitude vanishes if the overlap 
$\left\langle B_1 \bar{B}_2  | \Psi_{T_{bbb}} \right\rangle$
vanishes itself~\cite{Lea89}. Since there are no experimental data for the 
$\Omega_{bbb}$ mass and there is a wide variety of theoretical estimations (see
Table 1 of Ref.~\cite{Zha17}) it has to be calculated within the same scheme. 
For the binding energy of the $T_{bb}$ axial vector tetraquark obtained 
in Ref.~\cite{Fra17}, the $\Omega_{bbb}$ has a mass of 14.84\,GeV.
Thus, the $\Omega_{bbb} - \bar p$ threshold would lie at 15.78\,GeV, 
above the $T_{bbb}$ state. Let us note that even if the $\Omega_{bbb} - \bar p$ threshold would lie below the three $B-$meson
energy, the $T_{bbb}$ state will show up as a narrow resonance as recently discussed in Ref.~\cite{Gar18},
due to the negligible interaction between the $\Omega_{bbb}$ and the $\bar p$. The dynamics of this type of 
states would come controlled by the attraction in the three-body system and the channel made of almost non-interacting 
hadrons is mainly a tool for the detection. This is exactly the same situation 
observed in the case of the lower LHCb pentaquark $P^+_c(4380)$~\cite{Aai15} with a mass of
$4380\, \pm \, 8 \, \pm \, 29$ MeV, that it is seen to decay to the $J/\Psi - p$ channel 
with a width $\Gamma = 205 \, \pm \, 18 \, \pm \, 86$ MeV, while the phase space is of the order
of 345 MeV.

We have checked that the $T_{bbb}$ exotic state remains stable for the whole 
range of binding energies of the axial vector tetraquark $T_{bb}$
reported in the different theoretical studies~\cite{Fra17,Kar17,Eic17,Bic16,Vij09,Ric18,Luo17,Duc13,Cza18,Ade82}.
Thus, we have repeated the coupled-channel three-body calculation
for different binding energies of the axial vector tetraquark $T_{bb}$, starting
from the smallest binding of the order of 90\,MeV obtained in Ref.~\cite{Bic16}. 
The results are given in Table~\ref{tab2}. It can be seen that the three-meson 
bound state $T_{bbb}$ is comfortably stable for any of the binding energies of the 
axial vector tetraquark $T_{bb}$ reported in the literature. If the binding 
energy of the $T_{bb}$ state is reduced up to 50 MeV, the three-body system
would have a binding of the order of 23 MeV that would lie 
already 19 MeV above the lowest $BBB$ threshold, so that one does not expect
any kind of Borromean binding in this system. The situation is even worst
in the charm sector, because the vector-pseudoscalar meson mass difference
changes from 45 MeV in the bottom sector to 141 MeV in the charm sector,
so that the $DDD$ and $DDD^*$ thresholds would lie 
282 MeV and 141 MeV below the $DD^*D^*$ energy, respectively.

.\\
\begin{table}[t!]
\caption{Binding energy, in MeV, of the $T_{bbb}$ $(I)J^P=(1/2)2^-$
$B B^* B^* - B^* B^* B^*$ three-body system as a function
of the binding energy, in MeV, of the axial vector tetraquark $T_{bb}$.
The $T_{bbb}$ binding energy is calculated with respect to the 
lowest strong decay threshold: $m_B + 2 \, m_{B^*} - B(T_{bb})$.}
\begin{tabular}{|c|c|c|} 
\hline
$B(T_{bb})$ & $B(T_{bbb})$  \\ 
\hline\hline
180 & 90  \\
144 & 77  \\ 
117 & 57  \\
87  & 43  \\\hline																																					
\end{tabular}
\label{tab2}
\end{table}

\section{Summary.}
By solving exactly the Faddeev equations for the bound-state problem of three
mesons, we demonstrate that the current theoretical predictions pointing to 
the existence of a deeply-bound doubly bottom axial vector tetraquark lead to the 
likelihood of a bound state of three $B$ mesons.
We find that the $BB^*B^*-B^*B^*B^*$ state with $(I)J^P=(1/2)2^-$, $T_{bbb}$, is 
about 90\,MeV below any possible three $B$-meson threshold
for the standard binding of the recently reported axial vector doubly bottom tetraquark, $T_{bb}$.
It is important to note, as we have explained above, that this is the only three-body 
channel bringing together all necessary conditions about the two-body subsystems that allow
to maximize the binding of the three-body system. In other words, this 
unconventional form of a three-body hadron
is unique. The experimental search of these tetraquark, $T_{bb}$, and hexaquark, $T_{bbb}$, structures is a 
challenge well worth pursuing, because they are the first manifestly exotic
hadrons stable under strong and electromagnetic interaction.

It is appealing that the stability of such hexaquark state with respect 
the lowest tetraquark-meson threshold was already anticipated in the 
exploratory study of Ref.~\cite{Vij12} within a  quark string model.
Let us finally note that our discussion above could be extended to the charm sector, 
where the two-body bound state would lie close to threshold~\cite{Ike14,Jan04}.
However, as we have noted above, going from the bottom to the charm sector there is a factor 3 in the mass 
difference between pseudoscalar and vector mesons, what makes the coupled-channel effect much less 
important in the charm case than in the bottom one. Thus, one does not expect
binding in the three-meson charm sector.

\section{acknowledgments}
This work has been partially funded by COFAA-IPN (M\'exico), 
by Ministerio de Econom\'\i a, Industria y Competitividad 
and EU FEDER under Contract No. FPA2016-77177-C2-2-P
and by Junta de Castilla y Le\'on under Contract No. SA041U16.

\end{document}